# Waveguide photon-number-resolving detectors for quantum photonic integrated circuits

D. Sahin[1,a*], A. Gaggero[2*], Z. Zhou[1], S. Jahanmirinejad[1], F. Mattioli[2], R. Leoni[2], J. Beetz[3], M. Lermer[3], M. Kamp[3], S. Höfling[3] and A. Fiore[1]

[1]COBRA Research Institute, Eindhoven University of Technology, PO Box 513, 5600 MB Eindhoven, The Netherlands
[2]Istituto di Fotonica e Nanotecnologie, CNR, Via Cineto Romano 42, 00156 Roma, Italy
[3]Technische Physik, Physikalisches Institut and Wilhelm Conrad Röntgen Research Center for Complex Material Systems, Universität Würzburg, Am Hubland, D-97074 Würzburg, Germany

ABSTRACT

Quantum photonic integration circuits are a promising approach to scalable quantum processing with photons. Waveguide single-photon-detectors (WSPDs) based on superconducting nanowires have been recently shown to be compatible with single-photon sources for a monolithic integration. While standard WSPDs offer single-photon sensitivity, more complex superconducting nanowire structures can be configured to have photon-number-resolving capability. In this work, we present waveguide photon-number-resolving detectors (WPNRDs) on GaAs/Al$_{0.75}$Ga$_{0.25}$As ridge waveguides based on a series connection of nanowires. The detection of 0-4 photons has been demonstrated with a four-wire WPNRD, having a single electrical read-out. A device quantum efficiency ~24 % is reported at 1310 nm for the TE polarization.

*Authors contributed equally.
[a] Electronic mail: d.sahin@tue.nl

It is essential to increase the functionality and the complexity of quantum optics experiments in order to extend our understanding of interacting quantum systems and to provide a route to quantum information processing and manipulation. That requires increasing the number of quantum bits (qubits) in the quantum optical network to few tens and beyond. It is challenging to implement such systems with bulk optics due to the extreme stability requirements, the complexity and size, and the losses that scale proportionally. Integrated quantum photonics [1] is addressing those formidable challenges by replacing bulk optics with a more compact and efficient integrated configuration. In order to realize such quantum photonic integrated circuits (QPICs), single-photon sources, passive circuit elements such as waveguides, couplers and phase shifters, and single-photon detectors are required to be integrated on a single photonic chip [1, 2]. In particular, waveguide-single photon detectors (WSPDs) have been demonstrated recently, based on superconducting nanowires [3-6] and transition edge sensors (TESs) [7]. The superconducting nanowire approach can provide low dark count rates, excellent timing resolution and short dead time [8] and benefits from the high modal absorption of the guided mode that allows unity absorptance with waveguide lengths of a few tens of micrometers. Whilst integrated single-photon detectors are powerful components for a QPIC, detectors providing photon-number resolution are important in quantum communication and linear-optics quantum computing [9]. Recently, there has been a considerable effort to realize photon-number-resolving detectors (PNRDs) for free-space coupling using TESs [10], charge integration photon detectors [11], silicon photomultipliers [12] and avalanche photodiodes (APDs) [13], as well as time-multiplexing using Si-APDs [14] and SSPDs [15], and spatial multiplexing with APDs [16] and SSPDs [17-20]. Up to date, only transition-edge sensor detectors (TESs) have been reported in a waveguide configuration [7, 21]. Nevertheless, TESs are thermal detectors therefore they are relatively slow and unsuited for high-speed quantum information processing. In this report, we demonstrate waveguide photon-number-resolving detectors, utilizing NbN superconducting nanowires, which provide high efficiency and short deadtime.

Figure 1(a) shows a schematic of a waveguide photon-number-resolving detector (WPNRD). The detector is based on four NbN superconducting nanowires on top of a GaAs/Al$_{0.75}$Ga$_{0.25}$As (0.35 µm/1.5 μm-thick) waveguide heterostructure. The nanowires represent distinct detecting elements sensing different parts of the same waveguide mode and the number of switching wires can be determined from the output voltage as described below. We simulated a 3.85 μm-wide and 350 nm-thick ridge GaAs waveguide etched by 260 nm on top of $Al_{0.75}Ga_{0.25}As$ cladding layer with a finite-element solver (Comsol Multiphysics). The wires are 5 nm thick and 100 nm wide with a spacing of 150 nm and a total length of 60 (2x30) μm (Fig. 1(a)). In the simulation, we consider a 100 nm-thick $SiO_x$ layer that is left on top of the NbN nanowires as a residue of the hydrogen silsesquioxane (HSQ) resist after the patterning. The structure is optimized for nearly-equal absorption for different wires along the lateral direction of the waveguide. The symmetric configuration with a wider waveguide than WSPDs [3] is appropriately engineered to alleviate the difference in the absorption of the guided light by the central and lateral wires, while maintaining the absorptance of the quasi-transverse electric (TE) and transverse magnetic (TM) modes high. Moreover, the design is tolerant to the variation of the etching depth between 250 and 300 nm. We calculated the total absorptance for the lowest-order TE and TM modes, with the respective modal absorption coefficients of $\alpha_{tot}^{TE} = 478$ cm$^{-1}$ and $\alpha_{tot}^{TM} = 654$ cm$^{-1}$ (assuming $n_{NbN} = 5.23 - 5.82i$ [22]). As depicted in Fig. 1(b) that allows 76% TE and 86% TM absorptance along a 30 μm-long waveguide. The modal absorption coefficient by only the two central wires $\alpha_{cent}^{TE} = 282$ cm$^{-1}$, $\alpha_{cent}^{TM} = 380$ cm$^{-1}$ is higher than the corresponding absorption by the two lateral wires $\alpha_{lat}^{TE} = 198$ cm$^{-1}$, $\alpha_{lat}^{TM} = 276$ cm$^{-1}$ for both polarizations due to the confinement profile of the mode (see the inset of Fig. 1(b)). The probability of absorption after propagating over a length L, easily derived as $P_{cent(lat)}(L) = \frac{\alpha_{cent(lat)}}{\alpha_{tot}}(1 - e^{-\alpha_{tot}L})$, is plotted for both TE and TM polarizations for the two central (circles) (lateral (diamonds)) wires in Fig. 1(b). The situation is analogous to an unbalanced N-port splitter [23] followed by single-photon detectors. The corresponding unbalance in detection probability does not significantly limit the fidelity of the PNR measurement, as discussed below.

The electrical structure of WPNRDs is based on the series connection of four wires, each shunted by a resistance (see Fig. 1(a)) [19]. The photon detection mechanism in each wire is the same as in SSPDs

[8]. The wire is biased with a current close to its critical current ($I_c$), and upon absorption of a single photon, a resistive region is formed across it. While in SSPDs the bias current ($I_b$) is diverted to the external load resistance, in these series-nanowire detectors $I_b$ is redirected to the resistance integrated in parallel to each wire, producing a voltage pulse. The inset of Fig. 2(b) depicts the equivalent electrical circuit (showing only two wires for simplicity). If several wires switch simultaneously, a voltage approximately proportional to the number of switching wires is read on the load resistance [19].

WPNRDs integrated on a GaAs waveguide are defined using five steps of direct-writing electron beam lithography. We use a high resolution Vistec EBPG 5HR system equipped with a field emission gun with acceleration voltage 100 kV. In the first step, Ti(10nm)/Au(60nm) electrical contact pads (patterned as a 50 Ω coplanar transmission line) and alignment marks are defined using a positive tone polymethyl methacrylate electronic resist, evaporation and lift-off. In the second step, we define additional Ti(5nm)/Au(20nm) pads by electron beam lithography on polymethyl methacrylate, evaporation and lift-off. These pads are needed to allow the electrical connection between the nanowires and the parallel resistances (light green-colored pads in the inset of Fig. 2(a)). In the third step, the 100 nm wide meandered nanowires are defined on a 140 nm thick HSQ mask using an e-beam process optimized for GaAs substrates. The pattern is then transferred to the NbN film with a ($CHF_3$+$SF_6$+Ar) reactive ion etching. The left inset of Fig. 2(a) shows a scanning electron microscope (SEM) image of nanowires. In the fourth step, we fabricate the Ti(10 nm)/AuPd(50 nm) resistances. Each resistance is 500 nm wide and 3.5 μm long corresponding to a design value of $R_p$= 49 Ω. The right inset in Fig. 2(a) shows a magnified micrograph of the resistances. In the next step, we define the 180 nm thick and 3.85 μm wide HSQ-mask for the waveguide patterning by carefully realigning this layer with the previous one. This layer also protects the Ti/Au pads and the Ti/AuPd resistances during the GaAs etching process. Successively, we etch 260 nm of the underlying GaAs layer with a $Cl_2$+Ar electron cyclotron resonance etching. Finally, to allow probing the pads, holes are opened in the HSQ layer using a PMMA mask and reactive ion etching in $CHF_3$ plasma. The fabricated detector is shown in the SEM image in Fig. 2(a).

The experiments are performed by end-fire coupling near-infrared light from a lensed fiber to the waveguide, using the waveguide probe set-up described in Ref. [3]. Fig. 2(b) shows a characteristic current-voltage (IV) curve of a four-wire WPNRD. A critical current of $I_c = 10$ µA is measured at the base cold-plate temperature T = 2.1 K. The linear slope observed in the IV curve after reaching $I_c$ is related to the series connection of the four resistances, $4 \times R_p = 152$ Ω (38 Ω/each).

The system quantum efficiency (SQE) is defined as the number of counts (after subtracting the dark counts) divided by the number of photons at the fiber input of the cryostat. The SQE is measured by using a continuous-wave laser attenuated to the single photon level at 1310 nm and reaches 4% and 3.3% in the TE and TM polarizations, respectively. Fig. 3 shows the device quantum efficiency (DQE) of a WPNRD, defined as the number of photocounts divided by the number of photons coupled in the waveguide. The DQE reaches to 24±2 % for TE and 22±1 % for TM polarization at a bias current $I_b = 9.3$ µA and has been determined from the measured SQE and the coupling efficiency (η) of the photons from the fiber into the waveguide, $\eta_{TE} = 17\pm1$ % and $\eta_{TM} = 14.8\pm0.6$ % (SQE= DQExη). The value of η is approximately determined from the spectral average of the Fabry-Perot (FP) fringes measured on four, nominally identical waveguides (with no wires on top) by using a tunable laser around 1310 nm and its error bar is defined as the standard deviation among the four waveguides. For the TM polarization, this value of η corresponds well to the one determined ($\eta_{TM} = 14\pm1$ %) from the fringe contrast [3]. For the TE polarization, coupling to multiple lateral modes produces a complex fringe pattern, motivating our use of the spectral average. To date, this is the highest DQE reported for superconducting nanowire detectors with a single electrical output proportional to the photon number. The non-unity QE is attributed to the following reasons: 1- The absorptance of the 30 µm-long waveguides is calculated as 76% and 85% for the TE and TM polarizations, respectively. Longer wires may allow a higher DQE. 2- The deposition of very uniform NbN films is relatively difficult on GaAs [24] compared to the traditional substrates $Al_2O_3$ [8, 18] and MgO [25]. Whilst the sputtering requires high temperature to promote the surface diffusion of the sputtered particles and obtain a high quality film, the GaAs surface starts to become rough above 350 ºC [24]. Therefore, the film quality might also play a role in the quantum efficiency. We also observe a change in the ratio of the TE and TM efficiencies at low bias current, which seems to indicate a polarization dependent internal quantum efficiency (probability of detection once a photon is absorbed), as previously observed [22].

The temporal response of the WPNRD is probed with a TE polarized pulsed laser-diode (10 MHz) at 1310 nm using a sampling oscilloscope with the detector biased at $I_b$ = 8.8 µA. A photoresponse pulse corresponding to four-photon absorption is shown in the inset of Fig. 3. After performing a moving average over 10 data points (green line), a 1/e decay time of $\tau_{1/e}$ = 6.2 ns is calculated. That value agrees well with the value of $\tau_{1/e}$ = 5.6 ns obtained from the simulation using the electro-thermal model (red line) [19]. This corresponds to an estimated maximum count rate of > 50 MHz.

In order to show the proof of PNR capability, the device is tested under illumination with a pulsed laser diode (~100 ps pulse width, 2 MHz repetition rate), whose photon number distribution is described by Poissonian statistics, using a sampling oscilloscope after amplification by three amplifiers with a total gain of 43 dB. Fig 4(a) shows an example of a photoresponse of the detector in TE polarization for a photon flux of 12 photons/pulse in the waveguide, corresponding to an average number of detected photons $\mu_{av} \approx 2.3$ per pulse at $I_b$= 8.8 µA, with a DQE of 19 %. Five distinct detection levels in the figure correspond to the detection of 0-4 photons. The slow rise time of the photoresponse is due to the low-pass filter (DC-80 MHz) added to the circuit to remove the high frequency noise. After measuring the count rate at a fixed bias current, $I_b$ = 8.8 µA, as a function of the threshold voltage ($V_{th}$) of a frequency counter at different powers (12 MHz repetition rate, TE polarization), the plateaus corresponding to the different photon levels are determined. By setting the threshold levels in the counter according to the different photon levels, the detection probability relative to ≥1- (red), ≥2- (green), ≥3- (blue) and ≥4- (purple) photon absorption events is measured as a function of the power in the waveguide and plotted in Fig. 4(b). The results are in a good agreement with the expected detection probability $P(n|\mu) \propto \mu^n$ for a Poissonian source in the regime where detected average photon number µ is µ<<1, as shown by the $\mu^x$ fits (black lines) in Fig. 4(a) for each photon level. The inset in Fig. 4(b) shows the peak amplitudes ($V_{out}$) as a function of the detected photon numbers, together with a linear fit, showing the excellent linearity of the output voltage. The error bars represent the full-width-half-maximum (FWHM) of each peak which is nearly independent of the photon number, showing only about 20% increase from 0- to 4- photon level, and similar excess noise as observed in the first demonstration of a series-nanowire PNRD [20].

The fidelity (a measure of how precisely a PNRD can reconstruct the photon number) of WPNRDs is potentially affected by five factors: 1- limited efficiency, 2- limited number of wires, 3- the different absorption by the central and lateral wires, 4- signal-to-noise (S/N) ratio and 5- crosstalk (spurious switching of a wire after photon absorption in an adjacent one). According to our previous study on closely-packed wires in a similar configuration [26], crosstalk is negligible. We will evaluate the limitation in fidelity introduced by the other four factors for the case of detecting two photons in our 4-wire WPNRD [18]. Due to the limited efficiency (DQE = 0.24), the calculated probability of detecting two photons propagating in the waveguide is $P(2|2) = 0.058$. In a 4-wire WPNRD with unity efficiency and equal absorption probability on each wire, $P(2|2) = 0.75$ due to the probability that two photons are absorbed in the same wire. In our waveguide design with unbalanced absorptance in the central and the lateral wires, $P(2|2)$ would be slightly reduced to 0.74. Finally, the fidelity related to the overlap between the different photon levels (limited S/N ratio) is 0.97. We conclude that the fidelity in the present device is mainly limited by the efficiency [18] and could be increased to 0.74 by increasing the length and the internal efficiency. Further improvements require an increase in the number of wires and a more uniform absorption probability.

In conclusion, we have demonstrated WPNRDs based on NbN superconducting nanowires on a GaAs ridge waveguide. The detectors can resolve up to four photons and show device quantum efficiencies of 24% and 22% at 1310 nm for TE and TM polarized input light with an estimated maximum count rate of >50 MHz. The efficiency can be maximized by further optimizing the film quality and the fabrication process. These WPNRDs represent a substantial step towards the integration of highly-functional detectors in quantum photonic circuits.

Acknowledgements: This work was supported by Dutch Technology Foundation STW, applied science division of NWO, the Technology Program of the Ministry of Economic Affair, and by the European Commission through FP7 QUANTIP (Contract No. 244026).

Captions:

Fig. 1: (a) Schematic of a waveguide photon-number-resolving detector (WPNRD) consisting of four wires in series with a resistance ($R_p$) in parallel to each wire (contact pads are not shown). (b) Calculated absorptance of a WPNRD for TM (red, dashed line and empty symbols) and TE (black, continuous line and filled symbols) polarizations. The absorptance is calculated for the four wires (lines), the two central (circles) and the two lateral wires (diamonds). Inset: Contour plot of the electric field for the fundamental quasi-TE mode at 1300 nm.

Fig. 2: (a) Scanning electron microscope image of a WPNRD. Inset on the upper left: a blow-up image of the four wires before the waveguide etching step, where the wires have been colored for clarity. Inset on the upper right: a close-up, false-colored image of four AuPd parallel resistances (4x$R_p$). The scale bar of both the insets is 500 nm. (b) IV characteristic of a four-element WPNRD. The inset shows the equivalent circuit of the series connected nanowires (modeled with a normal resistance ($R_n$) and an inductance ($L_k$)), each shunted by an integrated resistance ($R_p$).

Fig. 3: Device quantum efficiency (device QE) of a WPNRD measured with TE and TM-polarized CW light at 1310 nm. Inset: Photoresponse pulse when four photons are detected. The green curve is the moving average of 10 data points showing a decay time of $\tau_d$= 6.2 ns and the red curve is the calculation from an electro-thermal simulation [12] after correcting for the filtering effect of the amplifiers (20 MHz - 6 GHz) when four photons are detected, giving a decay time of $\tau_d$= 5.6 ns.

Fig. 4: (a) An oscilloscope persistence map for a photon flux of 12 photons/pulse in the waveguide (3.7 pW average power), and corresponding measured (dark blue) pulse height distribution of one- to four-photon detection events (laser repetition rate of 2 MHz). A time

window of 50 ps (dark-yellow rectangle) around the voltage peak is used to make a histogram as shown on the left axis (black line) with the corresponding multi-Gaussian fit (dark yellow line, area underneath filled with light blue). Distinct levels are observed corresponding to the detection of 0-4 photons as indicated on the right axis. (b) Count rate measured with a pulsed laser (repetition rate of 12 MHz), corresponding to different photon counting levels: 1-photon (red), 2-photon (blue), 3-photon (green) and 4-photon (purple) and power-law fitting (black lines). The measurements in both (a) and (b) are done at $I_b$= 8.8 µA with a pulsed diode-laser in the TE polarization at 1310 nm. Inset: The signal amplitude as a function of the detected photon number, together with a linear fit (red dashed line). The black dots represent the peak voltage and the bars correspond to the full-width-half-maximum (FWHM) of each peak.

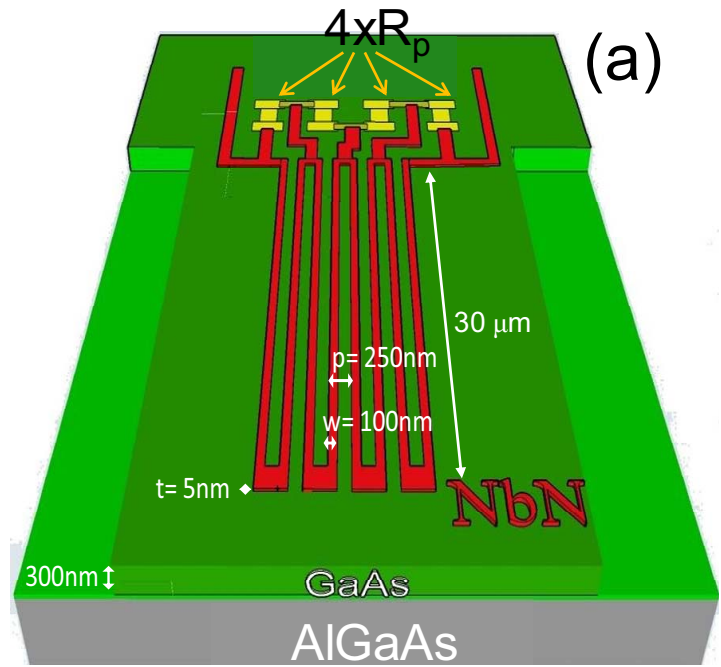
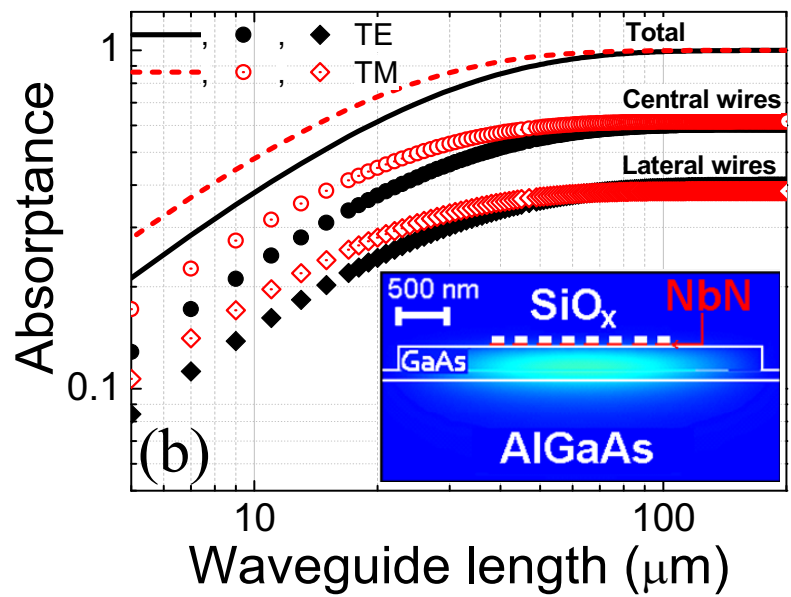

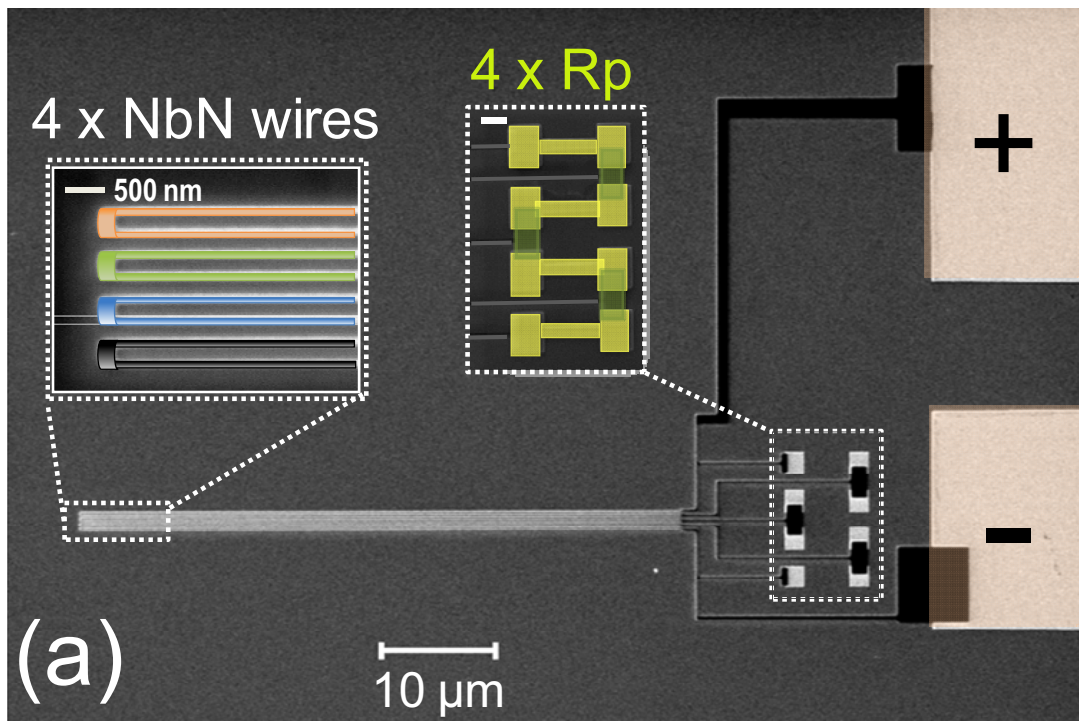
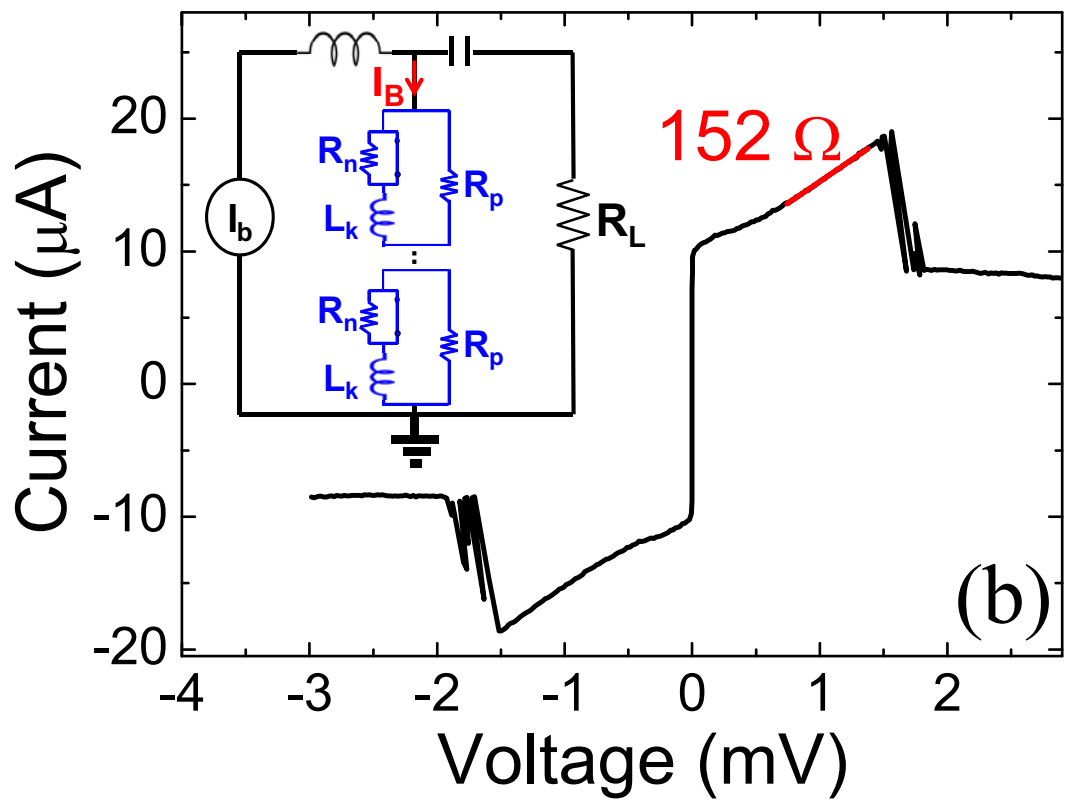

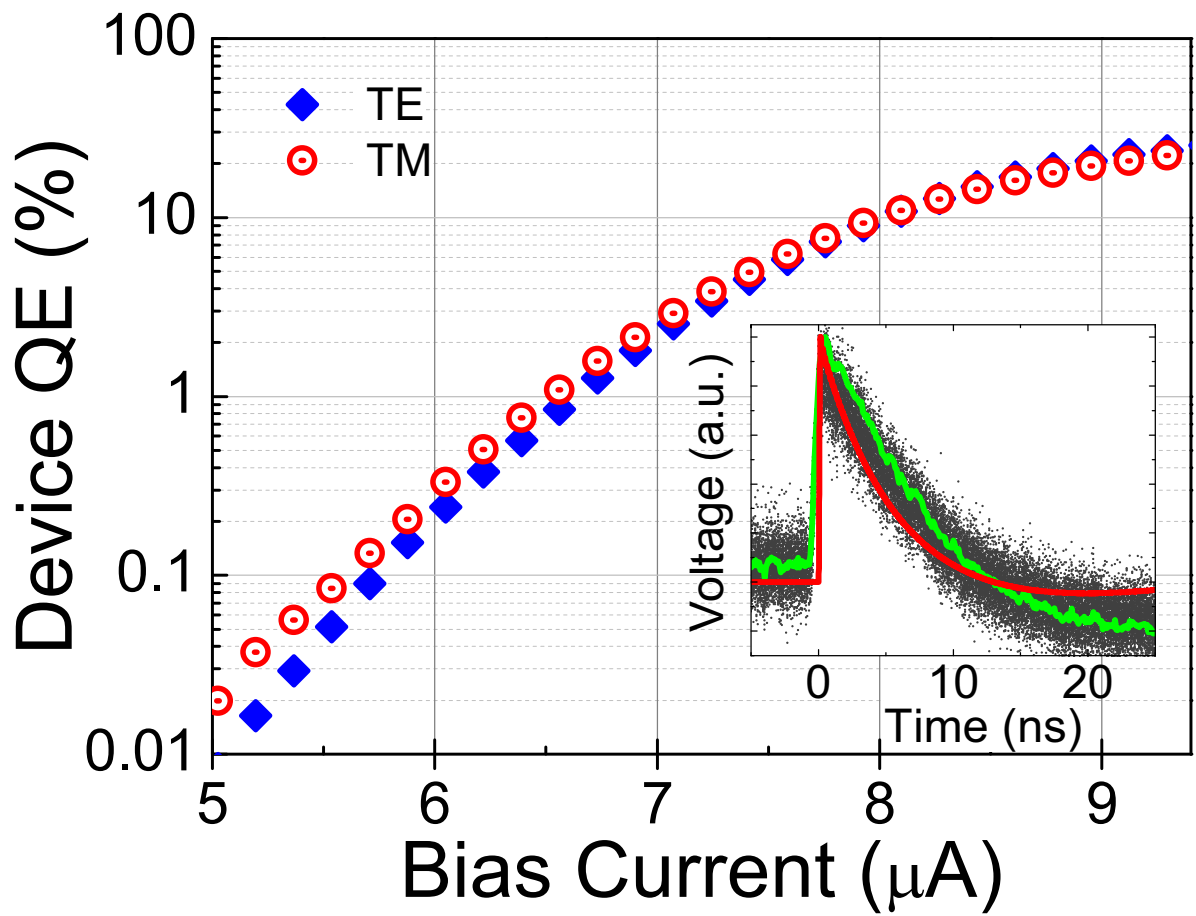

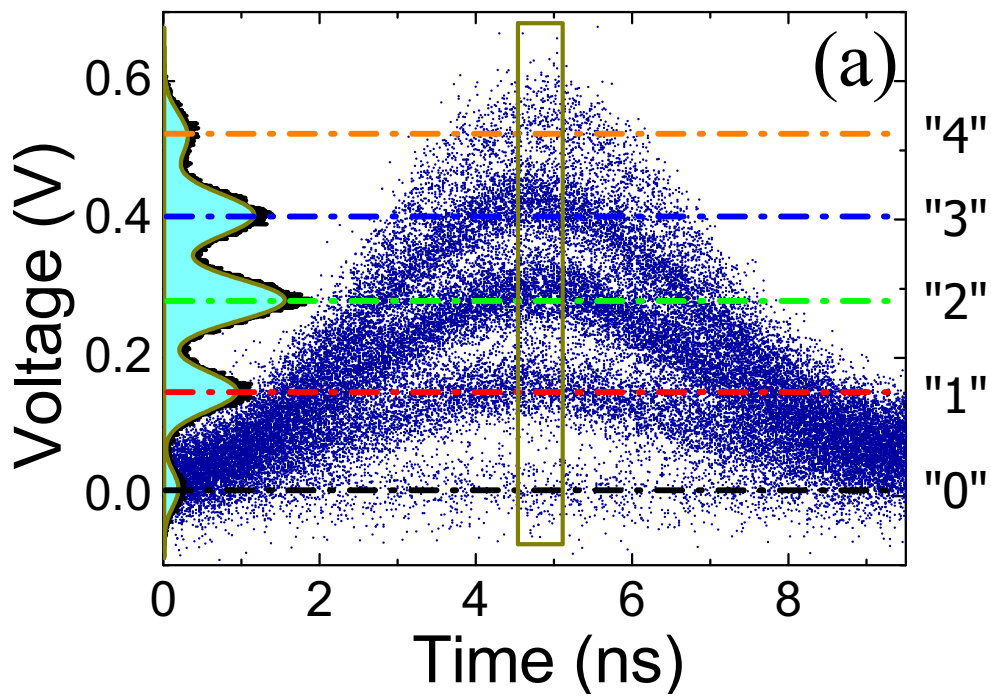
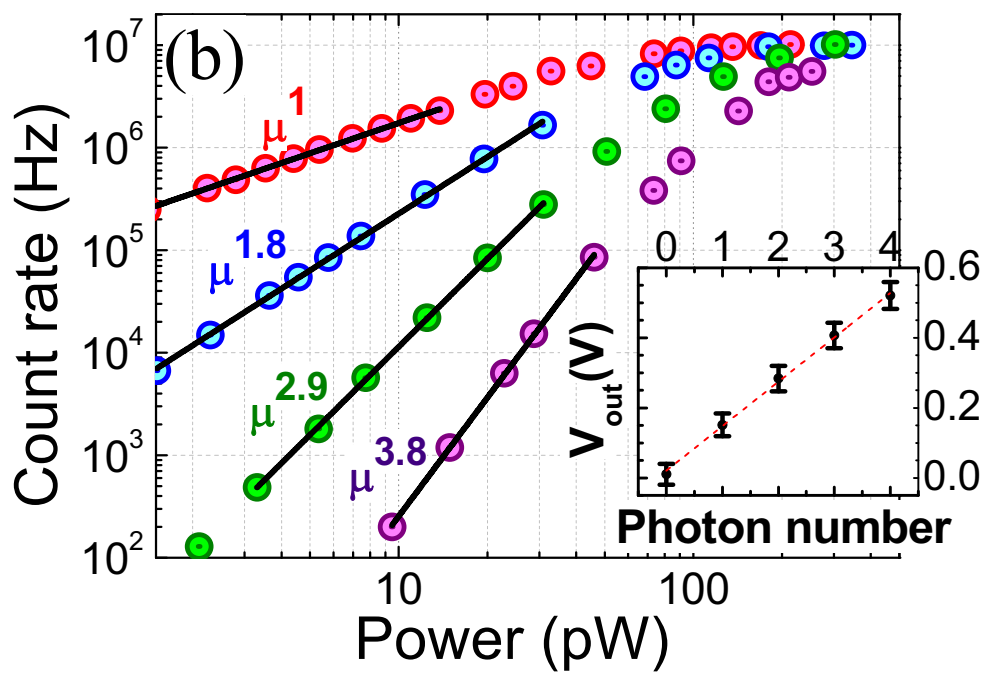